# "ChatGPT, a Friend or Foe for Education?" Analyzing the User's Perspectives on the Latest AI Chatbot Via Reddit

Forhan Bin Emdad[a], Benhur Ravuri[a], Lateef Ayinde[a], and Mohammad Ishtiaque Rahman[b]

[a]Florida State University, United States

[b]University of Maryland, Baltimore County, United States

fe21a@fsu.edu, bravuri@fsu.edu, layinde@fsu.edu, mrahman5@umbc.edu

## ABSTRACT

Latest developments in Artificial Intelligence (AI) and big data gave rise to Artificial Intelligent agents like Open AI's ChatGPT, which has recently become the fastest-growing application since Facebook and WhatsApp. ChatGPT has demonstrated its ability to impact students' classroom learning experience and exam outcomes. However, there is evidence that ChatGPT provides biased and erroneous information, yet students use ChatGPT in academic tasks. Therefore, an accurate understanding of ChatGPT user perception is crucial. This study has analyzed 247 Reddit top posts related to the educational use of ChatGPT from a prominent subreddit called "ChatGPT" for user perception analysis. Descriptive statistics, sentiment analysis using NLP techniques, and LDA topic modeling were used for analysis to gather a contextual understanding of the data. Results show that the majority of the users took a neutral viewpoint. However, there was more positive perception than negative regarding the usefulness of ChatGPT in education.

## KEYWORDS

social media, open-source software, user perception, artificial intelligence in education (AIEd), natural language processing, chatbots, ChatGPT

## INTRODUCTION

ChatGPT is the newest addition to the Artificial Intelligence (AI) family. ChatGPT, a large language conversation AI model, has over 100 million users by June 2023 (Ruby, 2023) of whom around 13.36\% are from the United States. The evolution of computer and information communication technologies has led to the growth of AI (Chen et al., 2020). Coppin (2004) describes AI as machines that can adjust and accommodate new situations and emerging technologies, solve new problems, answer questions, and perform many other tasks that require some kind of intelligence equivalent to what humans possess (Coppin, 2004). Artificial Intelligence (AI) has been applied to numerous domains, such as voice and visual recognition, decision-making, natural language processing, software application translation, embedded computer control systems, eye tracking system(Jamal et al., 2023), communication system (Moyen-ul-Ahsan et al., 2015) and robots (Chen et al., 2020). For instance, NLP models can train on a large number of clinical documents and generate knowledge that can assist experts in their clinical decision-making process (Jamal & Wimmer, 2023; Li et al., 2020).

Artificial Intelligence in Education (AIEd) is an emerging field that focuses on improving instructional design and student learning with the help of Artificial Intelligence (Xu & Ouyang, 2022). However, this research area has been well-established and has existed for over 30 years, focusing mainly on computers and education. Hwang et. al (2020) forecasted that future AI education research will largely center on implementing AI in education and methods to promote and teach AI knowledge at all levels of education (Hwang et al., 2020). Zawacki-Richter et al. (2019) established four potential applications in the field of education 1) Adaptive systems and personalization, 2) Assessment and Evaluation, 3) Profiling and prediction, and 4) Intelligent tutoring systems (Zawacki Richter et al., 2019). One aspect of the intelligent tutoring system in AIEd is 'Conversational AI' (Gao et al., 2018). From an educational perspective, conversational AI can perform well as an intelligent tutor assisting students with required study material, providing timely feedback, providing 24/7 guidance, and more (Mageira et al., 2022).

Large language models (LLMs) based AI agents and AI chatbots have recently grabbed global attention. LLMs are AI models which are trained with huge amounts of text data to complete several natural language processing (NLP) related tasks such as human communication, translation, information search, and text summarization (Ahn, 2023). ChatGPT is one of the latest developments in conversational AI (Ayinde et al., 2023; Taecharungroj, 2023). Though AI agents are yet to be human-like, AI agents such as ChatGPT can produce human-like output (Rudolph et al., 2023). Therefore, there is an ongoing debate among educators and other experts on the applications of conversational AI. Few experts believe that allowing AI agents can result in cheating (Susnjak, 2022) and others argue that accepting these AI agents will benefit them; for instance, AI agents can perform as teaching assistants (Qadir, 2022).

Despite differing opinions among the experts about AI applications, researchers emphasized on users' perspectives for evaluating perceived user satisfaction (Wang et al., 2022). AI developers can enhance the features of the AI agents and identify more improvement schemes for dissatisfied groups by utilizing user perception analysis methods. Therefore, we need to understand the users' views and perspectives as well as understand the stakeholder's (students, teachers, TAs, etc.) perceptions to make the ChatBot more efficient. Social media platform like Reddit allows users of different domains to gather, discuss, and share their valuable opinion anonymously in a specific AI agent-related community. On the contrary, Twitter reflects an individual's opinion without rebuttal from similar community members. Moreover, a robust social media data evaluation technique involving both human evaluation and AI-generated topic modeling is required to understand the context of the posts effectively. Hence, we used the Reddit platform to address the existing research gaps to explore the following research question (RQ) in our study:

RQ: What are the online audience's perceptions toward using conversational AI agents like ChatGPT from an educational context?

**RELATED WORKS**

Multiple studies tested the performance of an LLM (ChatGPT) in the United States Medical Licensing Exam, and the LLM could pass the test successfully (Gilson et al., 2023; Kung et al., 2023a). The authors suggest that these LLMs can assist in medical education and decision-making (Gilson et al., 2023; Kung et al., 2023a). In their study, Jiao et al. (2023) evaluated whether ChatGPT has an efficient translator. The LLM was tasked to translate a European language; the author suggests that ChatGPT's performance was equivalent to Google Translate (Jiao et al., 2023). Another study indicated that ChatGPT could assist with financial research (Dowling & Lucey, 2023). Two studies mentioned that ChatGPT saves time in the information search and retrieval process (Arif et al., 2023). Fijačko et al. (2023) used ChatGPT to pass the life support exams with satisfactory results (Fijačko et al., 2023). Choi et al. (2023) found that ChatGPT's performance in a law school exam was promising and they advised using ChatGPT as an assistant tool for legal writing (Choi et al., 2023).

However, three studies discussed the hallucination effect, which tends to happen when the users provide insufficient data to the LLM (Alkaissi & McFarlane, 2023). Arif et al. (2023) also mentioned that ChatGPT may lack critical thinking and clinical reasoning (Arif et al., 2023). Similarly, Cascella et al. (2023) argued that ChatGPT needs to be fine-tuned with medical-related data to answer medical-related questions as ChatGPT lacks medical knowledge and understanding of the context to answer complex clinical questions (Cascella et al., 2023; Emdad, Tian, et al., 2023). Finally, Kasneci et al. (2023) talked about bias and ethical issues in the output generated by ChatGPT (Emdad, Ho, et al., 2023; Kasneci et al., 2023).

There are few existing publications analyzing the perception of ChatGPT in the academic as well as in other significant domains. Firat (2023) identified the potential opportunities and challenges of ChatGPT in educational institutions by analyzing the perception of students and scholars by conducting a qualitative study (Firat, 2023). Praveen and Vajrabol (2023) analyzed the perception of healthcare researchers using BERT and topic modeling with Twitter data (Praveen & Vajrobol, 2023). Similarly, A survey conducted by Chan and Hu

(2023) determined students' perception of using ChatGPT for higher education was positive and had potential (Chan & Hu, 2023). However, the previous studies did not evaluate the perception of any community of social media containing all the stakeholders of education. Moreover, studies only implemented either a quantitative survey design with simple statistics models with human evaluation or social media data analysis without human evaluation.

**METHOD**

In this study, we searched for the communities relevant to education and ChatGPT in the Reddit social networking forum. We searched with the keyword "ChatGPT in education" on Reddit and found the "r/ChatGPT" subreddit with the highest number of members (539k) till March of 2023. This subreddit group was established on December 1, 2022, immediately after the release of ChatGPT.

**Data collection and preprocessing**

There are several filters on Reddit, i.e., top, hot, new, and rising posts. The research team targeted the top 1000 posts (dated between December 1, 2022, to March 15, 2023) that produced the most relevant information. The focus was on finding data regarding ChatGPT in education. We used PRAW, a well-known Python API, for collecting Reddit data. Next, the research team created a data dictionary consisting of the title and scores (calculated by the likes and comments from high karma users), upvotes received, URL, number of comments, post creation date, and description of the posts. Following the review process of removing the inconsistent data, the team ended up with 247 posts. Figure 1 shows that the number of posts is having an upward trend with a consistent increase.

**Figure 1**

*Number of Posts vs Time in Months for ChatGPT Subreddit (Dec 2022 - Mar 2023)*

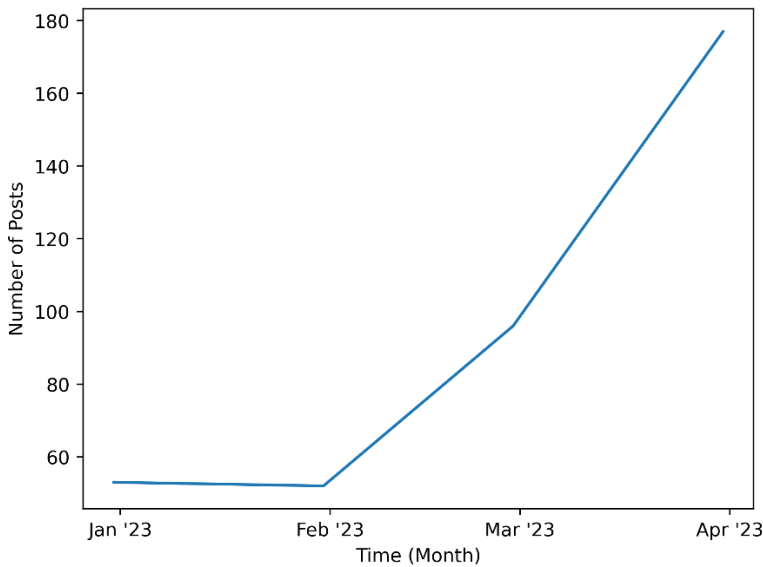

**Data analysis**

We conducted the study with a robust approach including three analysis steps. Firstly, we manually scrutinized the posts, and categorized them into two categories ('ChatGPT is useful' and 'ChatGPT is not useful') shown in Table 1 Then we calculated descriptive statistics with the combined reactions (derived from upvotes, scores, and the number of comments). Secondly, we conducted sentiment analysis with data tokenization and stop words removal and analyzed the data to identify the emotional perceptions of Reddit users using natural language processing (NLP) techniques. Sentiment analysis can provide a realistic view of the users' perception of AI agents (Garvey & Maskal, 2020).

Thirdly, we conducted Latent Dirichlet Allocation (LDA), a popular topic modeling approach, to identify the keywords in the top topics. Later, we compared the key terms of the topics with our manually identified keywords. Figure 2 provides an overview visualization of the study approach with each step described.

**Figure 2**

*Study Approach with Three Steps*

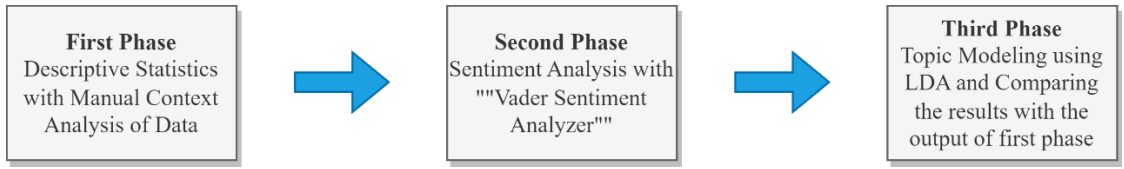

To achieve our objective, we filtered our Reddit data into two groups: 1) posts indicating the usefulness of ChatGPT in education and 2) posts indicating the adverse impacts of ChatGPT on education. We manually identified the connotation keywords. Table 1 provides the list of words used to categorize the posts.

**Table 1**

*Categorization of the Posts with the KeyWords*

| Categorization | Identified Key Words Indicating the Class |
|---|---|
| Useful | "best", "amazing", "good", "love", "Effectiveness", "empowers", "great", "new level", "time saver", "productivity", "booster", "usefulness", "unique", "better", "well", "Love", "Summarize", "success", "future", "helpful", "insights", "knowledge", "learn", "learning" |
| Not-useful | "dirty", "negative", "banning", "plagiarism", "censorship", "frustrating", "racist", |

| | "sexist", "beat", "lacking", "disallowed", "worse", "hack", "scary", "rejections", "garbage", "unnecessary", "unwanted", "Feminism-Hating", "violent", "falsifies citation", "destroying", "leaked", "re-think", "fake", "communist", "gulag", "war", "jailbreak", "nerfing", "computer winning", "ban", "odd filters", "Damn it" |
|---|---|

We implemented simple descriptive statistics on each group. The descriptive statistics included frequency (counts), mean, standard deviation, minimum, and maximum values. Next, we removed the inconsistent data to ensure accurate and reliable results. We then created a column named combined reactions combining scores, number of comments, and number of upvotes. Next, we sorted the data with combined reactions. Finally, we summarized the data with Python describe() functions to summarize important features of the dataset with numerical measures.

We used NLTK's "Vader Sentiment Analyzer" for the sentiment analysis. It assigns ranks to the title of the top posts with positive, negative, and neutral using Pythons' "vader lexicon" word library. Before this, we tokenized the data by breaking the stream into meaningful elements and removing stopwords using Python's spaCy library. Later, we estimated the sentiment with the get sentiment() function, which used the polarity score() function of the Vader analyzer to calculate the compound score. The compound score comes from adding up individual word scores. It ranges from -0.5 to 0.5, where -0.5 is negative, 0 is neutral, and 0.5 is positive.

In the topic modeling, we removed the stop words and stemmed the words with the "preprocess text()" function. Then, we created a dictionary with the "genism" module. Later, we identified topics and keywords using the LDA model and visualized the model output with the "pyLDAvis" library.

**RESULTS**

The descriptive statistics table (Table 2 shows that posts saying ChatGPT is useful in education have higher combined engagement metrics than posts saying it's not beneficial. There was a clear difference observed in mean, standard deviation (SD), count, and maximum scores except minimum was equal for both perceived useful and not useful posts about ChatGPT.

**Table 2**

*Descriptive Statistics of the Combined Engagement Metrics Regarding the Perception of ChatGPT in Education*

|  | Mean | SD | Count | Min | Max |
| --- | --- | --- | --- | --- | --- |
| Useful | 132.94736 | 273.62807 | 38 | 5 | 997 |
| Not-useful | 67.181818 | 127.28424 | 33 | 5 | 612 |

Similarly, observing the sentiment analysis output demonstrated in Table 3, we can see that the positive compound score is higher than the negative compound score, meaning the data contains more positive keywords and expressions than negative ones. The positive posts compound score is 0.44 and the negative post compound score is 0.41 (Here, the "+" and "-" sign only indicates positive and negative posts)

**Table 3**

*Number of Posts and Compound Scores for Neutral, Positive, and Negative Sentiment*

|  | Number of Posts | Compound Score |
| --- | --- | --- |
| Neutral  | 136 | 0.000190  |
| Positive | 69  | 0.4402590 |
| Negative | 42  | -0.410317 |

Figure 3 provides a clear visualization of the positive, negative, and neutral post count visualization in a bar chart derived from sentiment analysis with the Reddit data. We can visualize that number of positive posts is higher than the negative ones.

**Figure 3**

*Count of Neutral, Positive, and Negative Sentiment Presented with Bar Chart*

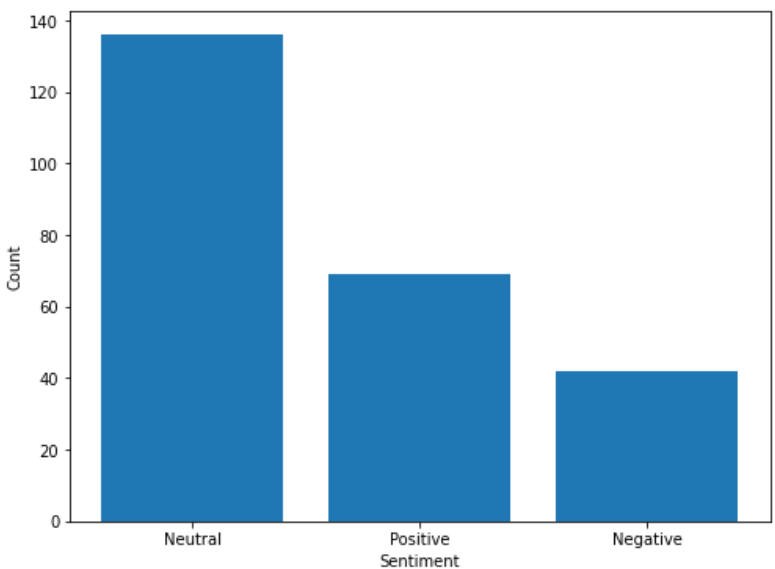

From implementing LDA, we identified the top 5 topics with the associated words or terms. Then, we compared the terms with our manually coded keywords and found similarities (highlighted in bold). Table 4 showed the top words associated with each topic. We can spot that new, learn, good, worse, and jailbreak are some of the important associated terms of the topics which are totally aligned with our manual coding of the data. In addition, we can observe that at least one similar word or term is present in each topic indicating fairness of the algorithm implementation.

Finally, Figure 4 shows a frequency visualization of the top 30 relevant terms in each topic as well as overall frequency. The blue bar indicates the overall frequency of the terms while the red bar indicates the frequency of the terms within the respective groups.

**Figure 4**

*Top-30 Most Relevant Terms for Each Topic*

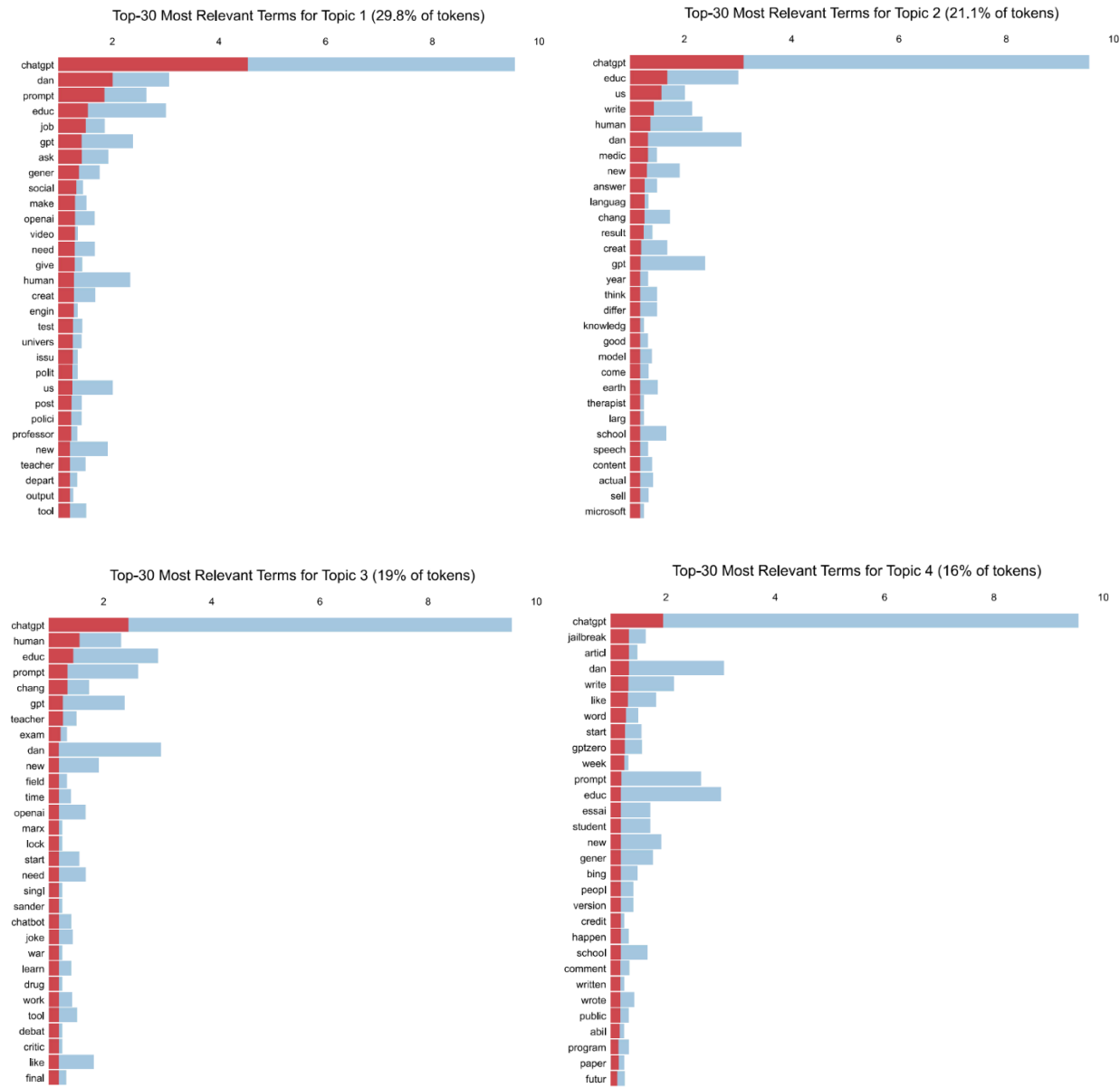

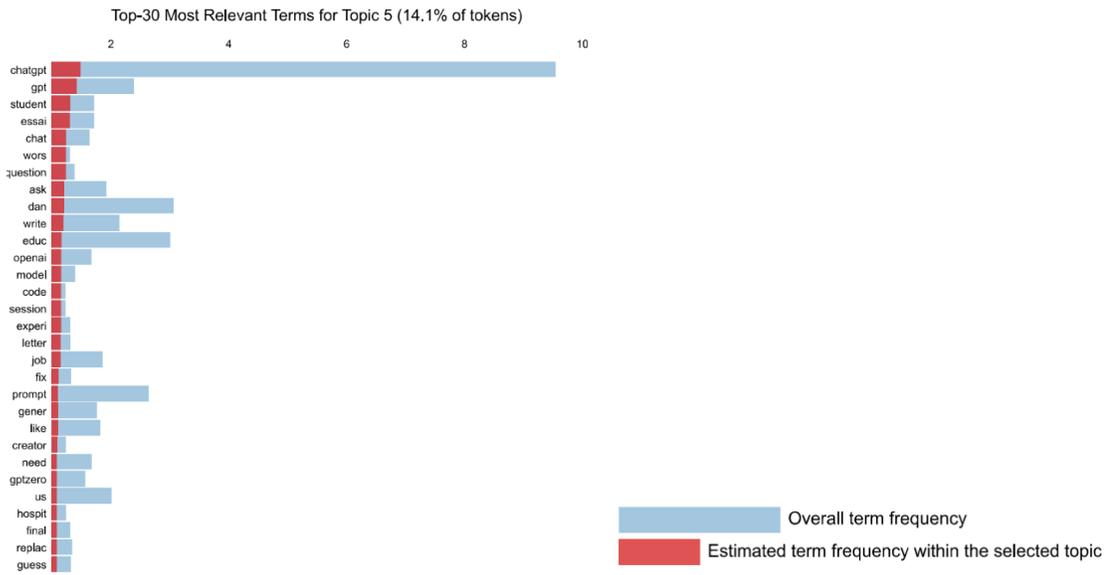

# Table 4

*Topics with 15 Associated words and Comparison with Manual Coding*

| Topics | Associated Words |
| --- | --- |
| Topic 1 | chatgpt, educ, us, write, human, dan, medic, **new**, answer, languag, chang, result, creat, gpt, year |
| Topic 2 | chatgpt, human, educ, prompt, chang, gpt, teacher, exam, dan, **new**, field, time, openai, marx, lock |
| Topic 3 | chatgpt, human, article, **jailbreak**, like, dan, school, **worse**, program, get, **good**, answer, accur, ask, new |
| Topic 4 | chatgpt, **jailbreak**, article, dan, write, like, word, start, gptzero, week, prompt, educ, essai, student, new |
| Topic 5 | chatgpt, gpt, student, essai, chat, **worse**, question, ask, dan, write, educ, openai, model, code, session |

## DISCUSSION

This research fills the gap identified in the literature regarding the perception of the usefulness of ChatGPT in education by using descriptive statistics, sentiment analysis, and LDA topic modeling on 247 individual posts. The study presents two primary conversation domains, 'usefulness' and 'not usefulness' of ChatGPT. "ChatGPT being useful" in education achieved a mean score of 132.95 and a count of 38 which is more than being "not useful" with a mean score of 67.18 and a count of 33. Through sentiment analysis, the study shows that most users are neutral (with a post count of 136) regarding their perception of using ChatGPT in education. However, our result also reveals that users have mixed perceptions about ChatGPT, spanning from neutral (136 posts), to positive (69 posts) and negative (42 posts). The compound score of the positive posts is 0.44 which is higher than the negative posts compound score (0.41).

Most scholars are neutral about the role of ChatGPT in education since it is a recently evolving tool. Several studies have been conducted on using ChatGPT for educational purposes (Kung et al., 2023b; Rudolph et al., 2023; Zhai, 2022). Some teachers and professors are skeptical about its use, as they do not know whether it is beneficial or harmful. According to our findings, ChatGPT is perceived as a useful and usable tool by users (Emdad & Koru, 2019). The results are in accordance with recent research by Baidoo-Anu, and Owusu Ansah (2023) (Baidoo-Anu & Owusu Ansah, 2023) and Sok and Heng (2023) (Sok & Heng, 2023), which concluded that ChatGPT is capable of being used in education as a tutoring tool, a grading system for essays, a language translator, an adaptive learning method. In support of our study, Qadir (2021) argued that ChatGPT could be a valuable educational tool (Qadir, 2022). The results of our study confirm those from Lund et al. (2023) concerning ChatGPT and its impact on academic and library communities (Lund & Wang, 2023). Moreover, Panda and Kaur (2023) demonstrate that ChatGPT improves user experience and reduces librarian workload by providing more accurate and personalized responses to user queries (Panda & Kaur, 2023). Despite the usefulness of ChatGPT in education, we also found that few users perceive ChatGPT as not useful. Tlili et al. (2023) identified ten scenarios in which ChatGPT might be involved in cheating, dishonesty, and untruthfulness (Tlili et al., 2023). According to Baidoo-Anu, and Owusu Ansah (2023), ChatGPT generates incorrect information, biases in trained data, interpretability and privacy concerns (Baidoo-Anu & Owusu Ansah, 2023; Feng et al., 2023). In addition, students could use it for plagiarism which may result in students not understanding complex tasks when over-reliance on technology (Islam & Islam, 2023).

Sentiment analysis of the social network data is not always accurate to identify the perception of users as the posts can be influenced by many surrounding factors, proper context, and can involve sarcasm. Although we tried manually reviewing the text data in our study to minimize this problem. However, there can still be bias in the manual analysis of this data.

**CONCLUSION**

This study recommends that policymakers, researchers, educators, and technology experts can take note of the findings of this study and consider ChatGPT as a great benefit to education. Educators and students should embrace technology to improve the teaching and learning process to improve education and support classroom learning. Conclusively, this study points out that ChatGPT in education is a helpful tool and the need for educators to rethink how they teach by using ChatGPT as a teaching tool. From a practical perspective, educators should incorporate the use of Chatbots into school curricula and upskill teachers in the use of this technology. Future research could use qualitative and quantitative methods to understand the online audience's perception of using generative AI agents in education.

**Supplementary information**

We used the Google Colab notebooks platform to implement our perception analysis codes. All the codes used for this study are available in author's github repository.

**Supplementary information**

**Funding** This research was not funded by any funding agency.

**Conflicts of interest** The author has no relevant financial or non-financial interests to disclose.